\documentclass{ws-procs9x6}

\begin{document}

\title{
\hfill{\small {\bf MKPH-T-04-15}}\\ \vspace*{.2cm}
Spin asymmetry and GDH sum rule for real and virtual photons for 
the deuteron\footnote{\uppercase{W}ork supported by \uppercase{D}eutsche 
\uppercase{F}orschungsgemeinschaft (\uppercase{SFB} 443).}
}
\author{H. Arenh\"ovel, A. Fix and M. Schwamb}
\address{Institut f\"ur Kernphysik, Johannes Gutenberg-Universit\"at, 
D-55099 Mainz,\\ Germany\\
E-mail: arenhoevel@kph.uni-mainz.de}

\maketitle

\vspace*{-.7cm}

\abstracts{An explicit evaluation of the spin asymmetry of 
the deuteron and the associated GDH sum rule is presented which includes  
disintegration, single and double pion and eta production. 
For the GDH integral a large cancellation is found 
between the disintegration channel and the meson production channels.
Furthermore, first results for the contribution of the disintegration 
channel to the generalized GDH integral at constant four-momentum transfer 
reveal a dominance of the isovector M1 transition to the 
$^1$S$_0$-state near threshold resulting in a negative contribution with 
a minimum around $Q^2\approx 0.2$~fm$^{-2}$ which is driven by the 
nucleon anomalous isovector magnetic moment.}
\vspace*{-1.2cm}

\section{Introduction}
The {\sc Gerasimov-Drell-Hearn} sum rule links 
the anomalous magnetic moment $\kappa$ of a particle to the integral over the 
energy weighted spin asymmetry of the absorption cross section 
with respect to circularly polarized photons and a polarized target
\begin{eqnarray}
I^{GDH}=\int_0^\infty \frac{d\omega}{\omega}
\Big(\sigma^P(\omega)-\sigma^A(\omega)\Big)
\,&=&\,\, 4\,\pi^2 \kappa^2\frac{e^2}{M^2}\,S\,,
\end{eqnarray}
where $S$ denotes the spin of the particle and $M$ its mass.
Obviously, for $\kappa \neq 0$ the particle possesses 
an internal structure. However, the opposite is not in general true. 
A particle having a vanishing or very small $\kappa$ need not be
pointlike or nearly pointlike. The deuteron is an instructive example
for such a case.

\section{GDH Sum Rule for the Deuteron\protect\cite{ArF04}}
The deuteron is an isoscalar with a very small 
$\kappa_d \,=\,-\,0.143$~n.m. and thus a very small sum rule value
$I^{GDH}_d\,=\, 0.65\, \mu\mbox{b}$. On the other hand it is well known, 
that the deuteron has quite an extended spatial structure due to its 
small binding energy. In fact, the small $\kappa$ arises from an almost 
complete cancellation of proton and neutron anomalous magnetic moments in 
the deuteron because of their parallel spin orientation. Thus
it is expected that also for the sum rule integral such a cancellation occurs.
As absorptive processes one distinguishes (i) 
photodisintegration $\gamma + d \, \rightarrow  \,n + p$
and (ii) meson production. The latter process is certainly dominated by 
quasi-free production on the nucleons, and
a simple estimate of its contribution to the sum rule is obtained by
summing incoherently neutron and proton GDH contributions 
$I^{GDH}_p + I^{GDH}_n\,=\, 438\, \mu\mbox{b}$, 
neglecting interference and other binding effects. In view of the tiny 
deuteron GDH value a negative contribution of almost equal size has to 
come from the photodisintegration channel. 

\begin{figure}[ht]
\vspace*{-.4cm}
\centerline{\epsfxsize=10cm\epsfbox{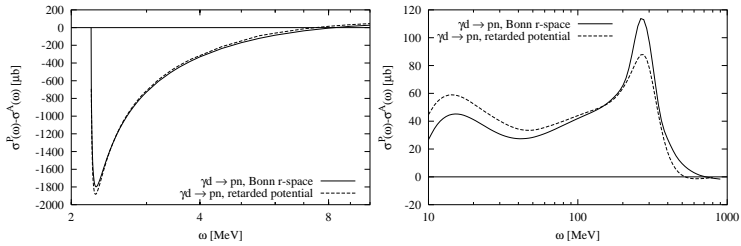}}
\vspace*{-.2cm}
\caption{Spin asymmetry of deuteron photodisintegration 
using (a) Bonn r-space potential\protect\cite{MaH87}+MEC+IC(IA)+RC and
(b) retarded potential + retarded $\pi$-MEC, 
$\Delta$-degrees in coupled channel, $\pi d$-channel 
+ RC\protect\cite{ScA01}. Left panel: low energy; right panel: high energy.}
\vspace*{-.5cm}
\label{spin_asy_np}
\end{figure}

Indeed, near threshold the spin 
asymmetry of photodisintegration is dominated by the 
isovector M1 transition to the 
$^1$S$_0$-state. Since it can only be reached for 
antiparallel photon and deuteron spins, a negative 
spin asymmetry arises which is quite huge 
as is shown in Fig.~\ref{spin_asy_np}. Besides an earlier 
evaluation\cite{ArK97} with static interactions and exchange currents 
(MEC), isobar configurations (IC) in impulse approximation (IA) 
and relativistic contributions (RC), we show in addition the result of a 
recent calculation\cite{ScA01} based on a retarded potential with retarded 
$\pi$-MEC, a coupled channel $N\Delta$-dynamics, 
and RC. 
It leads to a significant change of the spin
asymmetry and thus its GDH contribution.

\begin{figure}[ht]
\vspace*{-.5cm}
\centerline{\epsfxsize=8cm\epsfbox{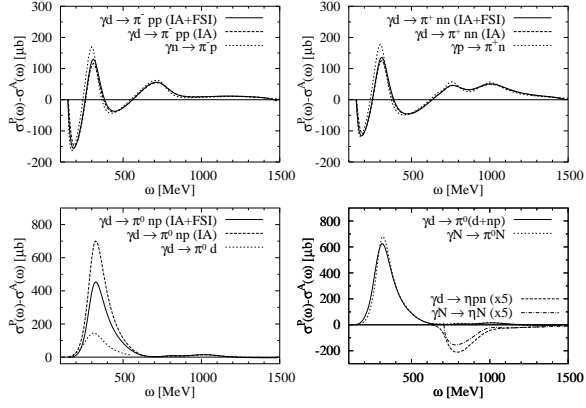}}
\vspace*{-.2cm}
\caption{Spin asymmetries of single pion and eta production on nucleon 
and deuteron for various charge channels. The results for the deuteron are in 
IA and with inclusion of FSI in final NN- and $\pi$N-subsystems. Upper panels: 
charged single pion production; lower left panel: coherent and incoherent 
$\pi^0$-production on deuteron; lower right panel: total 
$\pi^0$- and $\eta$-production on nucleon and deuteron. For $\pi^0$- and 
$\eta$-production on the nucleon the sum of spin asymmetries of 
neutron and proton is shown. The result for coherent $\pi^0$-production 
on the deuteron is taken from Ref.\protect\cite{ArK97}.}
\vspace*{-.2cm}
\label{spin_asy_piNN}
\end{figure}

An improved calculation of the spin asymmetries of single pion 
and eta production on the deuteron has been performed recently 
in which for the elementary production operator the MAID 
model\cite{MAID} has been used and 
final state interactions (FSI) are included completely in the  NN- and 
$\pi$N-subsystems. The results are shown in Fig.~\ref{spin_asy_piNN}.
For charged pion production the FSI effects are small, but quite 
sizeable for incoherent neutral pion production due to the 
non-orthogonality of the final state in impulse approximation. 
For comparison, the corresponding spin asymmetries for pion and eta 
production on the nucleon are also shown in Fig.~\ref{spin_asy_piNN}.
Significant differences between the spin asymmetries
of nucleon and deuteron are readily seen which certainly prevent one to
extract in a simple manner the neutron spin asymmetries from
deuteron data.

\begin{figure}[ht]
\vspace*{-.3cm}
\centerline{\epsfxsize=8cm\epsfbox{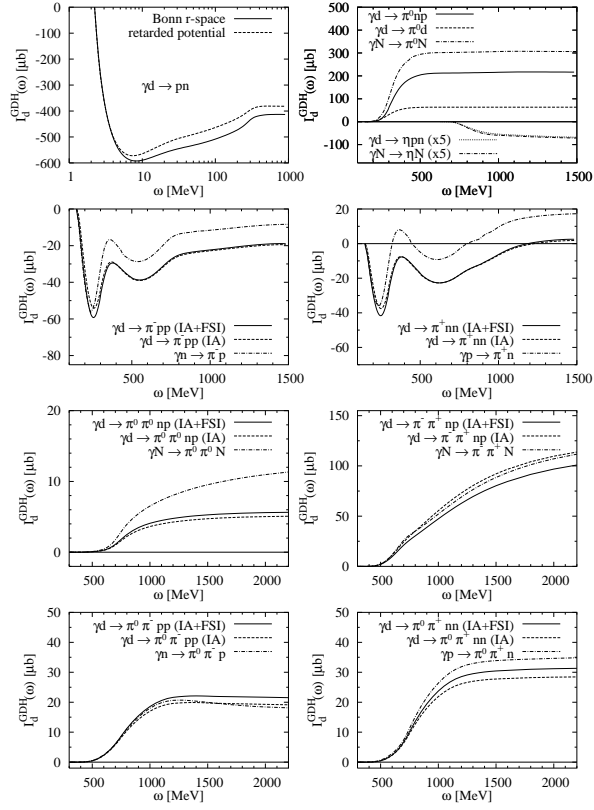}}
\vspace*{-.2cm}
\caption{Contribution of various channels to the finite GDH integral 
as function of the upper integration limit for deuteron disintegration, 
single and double pion and eta production on nucleon and deuteron. For 
the neutral charge channels $\pi^0$, $\eta$, $\pi^0\pi^0$, and $\pi^-\pi^+$, 
the nucleon integrals are the sum of neutron and proton integrals.}
\vspace*{-.4cm}
\label{int_gdh_all}
\end{figure}

\begin{table}[phtb]
\vspace*{-.2cm}
\tbl{Contributions of single $\pi$-production on nucleon and 
deuteron to finite GDH integral up to 1.5 GeV in $\mu$b.}
{\footnotesize
\begin{tabular}{@{}cccccc@{}}
\hline
{} &{} &{} &{} &{} &{}\\[-1.5ex]
 nucleon & $\pi^0p$ & $\pi^0n$ & $\pi^0(n+p)$ & $\pi^-p$  & $\pi^+n$ \\[1ex]
\hline
{} &{} &{} &{} &{} &{}\\[-1.5ex]
 & 159.10 & 147.34 & 306.44 & $-8.39$ & 17.28 \\[1ex]
\hline
{} &{} &{} &{} &{} &{}\\[-1.5ex]
deuteron &    $\pi^0d$ & $\pi^0np$ & $\pi^0(d+np)$ & $\pi^-pp$ & $\pi^+nn$ \\[1ex]
\hline
{} &{} &{} &{} &{} &{}\\[-1.5ex]
&  $63.26$ & 216.61 & 279.87 & $-18.94 $ &  $2.51$ \\[1ex]
\hline
\end{tabular}}
\label{tab1}
\vspace*{-.1cm}
\end{table}

Now we will turn to the explicit evaluation of the finite GDH integral
as defined by
\begin{eqnarray}
I^{GDH}(\omega)=\int_0^\omega \frac{d\omega'}{\omega'}
(\sigma^P(\omega')-\sigma^A(\omega'))\,.
\end{eqnarray}
The results for photodisintegration, single and double pion and eta 
production with and without FSI are exhibited in Fig.~\ref{int_gdh_all}
together with the corresponding nucleon values. The two-pion results are 
based on a recent evaluation using an effective Lagrangean approach 
analogous to Ref.\cite{GoT96}. 
It is obvious that convergence of the GDH integral is reached for 
$\gamma d\rightarrow np$ already at about 0.8 GeV and for neutral pion
production at 1.5 GeV, however, not completely for charged $\pi$-production. 
Also two-pion production has not reached convergence at 2.2~GeV, in 
particular for $\pi^-\pi^+$ production. One furthermore notes a 
sizeable reduction of the $np$ contribution from retardation 
in potential and $\pi$-MEC. For comparison, the finite GDH integrals 
for meson production on the nucleon are also shown in
Fig.~\ref{int_gdh_all} where one sees quite clearly 
significant differences between nucleon and deuteron results 
because of (i) Fermi motion and 
(ii) final state interactions. The results for the GDH integrals 
for single pion production on nucleon 
and deuteron, are listed in Table~\ref{tab1}. Also here one clearly notes 
significant differences between the nucleon and deuteron values.

\begin{table}[phtb]
\vspace*{-.1cm}
\tbl{Contributions of various channels to the finite
GDH integral (in $\mu$b), integrated up to 0.8~GeV for $\gamma d\rightarrow np$, 1.5~GeV
for single pion and eta production and 2.2~GeV for double pion production
on nucleon and deuteron.}
{\footnotesize
\begin{tabular}{@{}ccccccc@{}}
\hline
{} &{} &{}&{}&{}&{}&{} \\[-1.5ex]
 &  np & $\pi$ & $\pi\pi$ & $\eta$ & $\Sigma$ & sum rule \\[1ex]
\hline
{} &{} &{}&{}&{}&{}&{} \\[-1.5ex]
 n+p  &           &  315.33 & 175.95 & $-14.54$ & 476.74 & 437.94\\[1ex]
 d & $-381.52$ & $263.44$& 159.34 & $-13.95$ & $27.31$ & 0.65\\[1ex]
\hline
\end{tabular}}
\label{tab2}
\vspace*{-.3cm}
\end{table}
The GDH contributions of various channels to nucleon 
(sum of neutron and proton) and deuteron are listed in Table~\ref{tab2}.
The deuteron results include NN-FSI.
Compared to our earlier evaluation one now finds indeed for the deuteron 
a large cancellation between the various channels contributing to the 
GDH sum rule. This strong cancellation between the regions at low and high 
energies is a fascinating feature clearly demonstrating the decisive 
role of the pion as a manifestation of chiral symmetry governing 
strong interaction dynamics in these two different energy regions.
The cancellation constitutes also a challenge for any theoretical 
framework since 
it requires a unified consistent description of hadron and e.m.\ 
properties for both energy regions.

At the end of this section we would like to recall our concern with respect
to attempts to extract the neutron spin asymmetry from the one 
of the deuteron by subtracting the proton asymmetry. We will not repeat the
arguments which had been put forward by one of us\cite{Are01}. 
We only would like to emphasize that these arguments 
are confirmed by the results in Fig.~\ref{int_gdh_all} and in 
Table~\ref{tab2}. 

We only would like to add the following consideration: Assuming 
the validity of the sum rule and assuming that indeed the contribution 
from meson production on the deuteron would result in the sum of 
neutron and proton GDH sume rules (438 $\mu$b) than it would mean 
that the contribution of the photodisintegration channel has to be 
almost equal the negative value of this sum except for the deuteron
sum rule value, i.e. $-(437.94-0.65)~\mu$b. But the best theory at 
present yields only a converged value of $-381.52~\mu$b for this channel. 

\section{Generalized GDH Sum Rule for Virtual Photons\protect\cite{Are04}}

The spin asymmetry for real photons corresponds to the 
beam-target vector asymmetry
$A_{ed}^V$ of the general inclusive cross section of deuteron 
electrodisintegration\cite{LeT91} 
\begin{eqnarray}
\sigma (h, P^d_1 , P^d_2)=\sigma_0\,(1+P_1^d\,A_d^V+P_2^d\,A_d^T
     +h\,[A_e+P_1^d\,A_{ed}^V+P_2^d\,A_{ed}^T]),
\end{eqnarray}
where $\sigma_0$ denotes the unpolarized cross section, $h$ the electron 
polarization, and $P^d_1$ and $P^d_2$ deuteron vector and tensor polarization, 
respectively. The various asymmetries depend on the deuteron orientation
angles $\theta_d$ and $\phi_d$. 
For deuteron orientation parallel to the momentum 
transfer ${\vec {q}}$ the asymmetry $A_{ed}^V$ is determined by the 
transverse form factor $F^{\prime 10}_T$ which in turn at the 
photon point is related to the spin asymmetry for real photons\cite{LeT91}  
\begin{eqnarray}
\sigma_\gamma ^P(\omega^{lab})-\sigma_\gamma ^A(\omega^{lab})=
\frac{\sqrt{6}\,M_d}{{W_{np}\,q^{c.m.}}}F^{\prime 10}_T|_{Q^2=0},
\end{eqnarray}
where $W_{np}$ denotes the invariant mass of the $np$ system.
Therefore, we introduce as spin asymmetry for transverse virtual photons
\begin{eqnarray}
\sigma_{T,\gamma^*} ^P (\omega^{lab},Q^2)-
\sigma_{T,\gamma^*} ^A (\omega^{lab},Q^2)=
\frac{\sqrt{6}\,M_d}{{W_{np}\,q^{c.m.}}}F^{\prime 10}_T (\omega^{lab},Q^2).
\end{eqnarray}

\begin{figure}[ht]
\vspace*{-.7cm}
\centerline{\epsfxsize=9cm\epsfbox{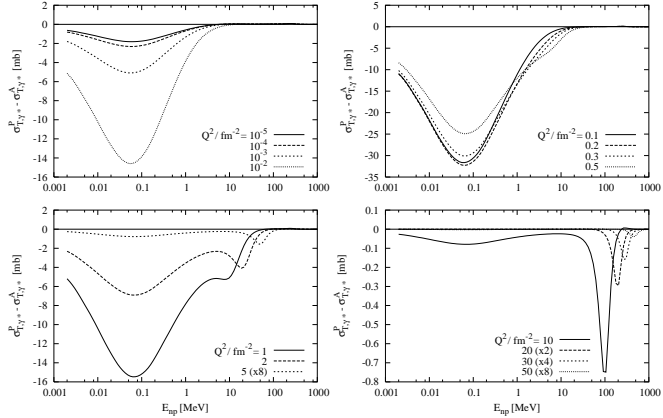}}
\vspace*{-.2cm}
\caption{Transverse spin asymmetry $\sigma_{T,\gamma^*}
^P-\sigma_{T,\gamma^*} ^A$ of $d(e,e')np$ as function of $E_{np}$ 
for various values of $Q^2$. Calculation based on 
Argonne $V_{18}$ potential\protect\cite{WiS95} including interaction 
and relativistic 
effects.}
\vspace*{-.3cm}
\label{fig_gen_gdh1}
\end{figure}
\begin{figure}[ht]
\vspace*{-.5cm}
\centerline{\epsfxsize=5.5cm\epsfbox{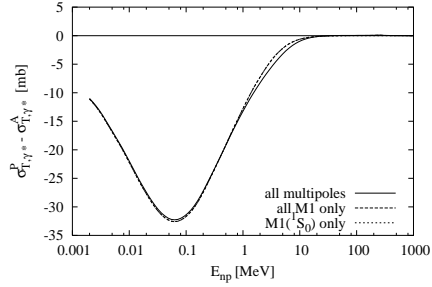}}
\vspace*{.1cm}
\caption{Comparison of the transverse spin asymmetry for the
M1 transition $^3$S$_1 \rightarrow$ $^1$S$_0$ alone to all M1 transitions 
and to all multipoles for $Q^2=0.2$~fm$^{-2}$.}
\label{compare_spin_asy_M1}
\vspace*{-.5cm}
\end{figure}

The transverse spin asymmetry has been evaluated for the 
electrodisintegration channel for various values 
of $Q^2=$~const with inclusion of MEC, IC and RC\cite{Are04}. 
The results are shown in 
Fig.~\ref{fig_gen_gdh1}. A detailed analysis\cite{Are04} has revealed that 
also for virtual photons the transverse spin asymmetry is dominated 
near threshold by the isovector M1
transition to the ``antibound'' $^1$S$_0$-state resulting in a large 
negative contribution which is deepest around $Q^2\approx 0.2$~fm$^{-2}$. 
The rapid fall-off with increasing $E_{np}$ ensures a good convergence. 
For higher $Q^2$ a negative peak at
quasi-free kinematics $E_{np}/$MeV$=10\, (q^{c.m.})^2/$fm$^{-2}$ emerges
from quasi-free scattering off neutron and proton which can
occur for antiparallel spin orientation only. However, its
amplitude decreases rapidly with growing $Q^2$.
The dominance of the M1 transition into the $^1$S$_0$ state near threshold
is demonstrated in Fig.~\ref{compare_spin_asy_M1}.

We now turn to the definition of the generalized GDH integral
\begin{eqnarray}
I_{\gamma^* d}^{GDH}(Q^2)=\sqrt{6}\int_{\omega_{th}^{lab}}^\infty 
\frac{d\omega^{lab}}{\omega^{lab}}\,\frac{M_d\,g(\omega^{lab},Q^2)}
{W_{np}\,q^{c.m.}}F_{T} ^{\prime 10},
\end{eqnarray}
where $F_{T} ^{\prime 10}=F_{T} ^{\prime 10}(E_{np},q^{c.m.})$ is
an implicit function of $\omega^{lab}$ and $Q^2$.
The factor $g(\omega^{lab},Q^2)$ appears because the generalization of 
the GDH integral is to a certain extent arbitrary. The only restrictions 
are (i) at the photon point $Q^2=0$ the condition $g(\omega^{lab},0)=1$,
and (ii) 
\begin{eqnarray}
\lim_{\omega^{lab}\rightarrow \infty}g(\omega^{lab},Q^2)|_{Q^2=const.}
<\infty \,.
\end{eqnarray}
As simplest extension we have choosen $g(\omega^{lab},Q^2)\equiv 1$.

\begin{figure}[ht]
\vspace*{-.3cm}
\centerline{\epsfxsize=5cm\epsfbox{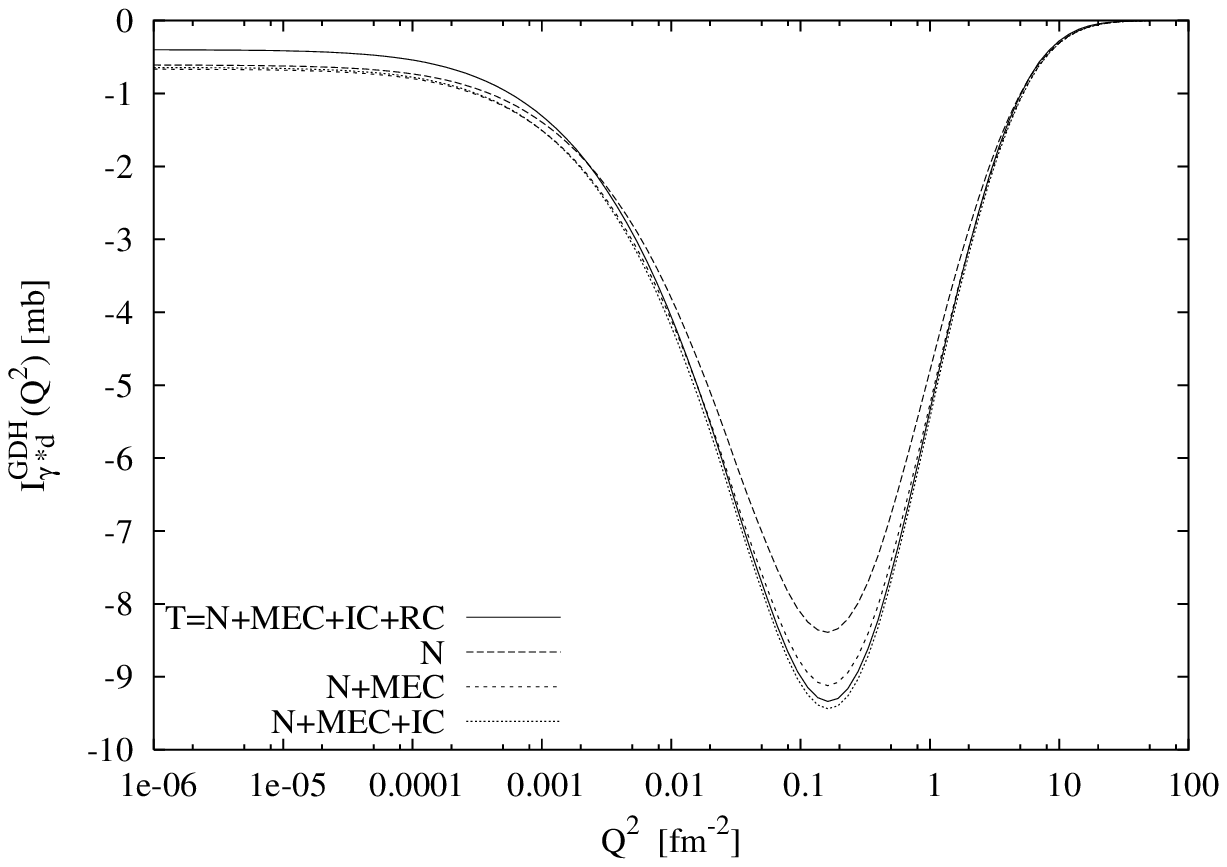}\epsfxsize=5cm\epsfbox{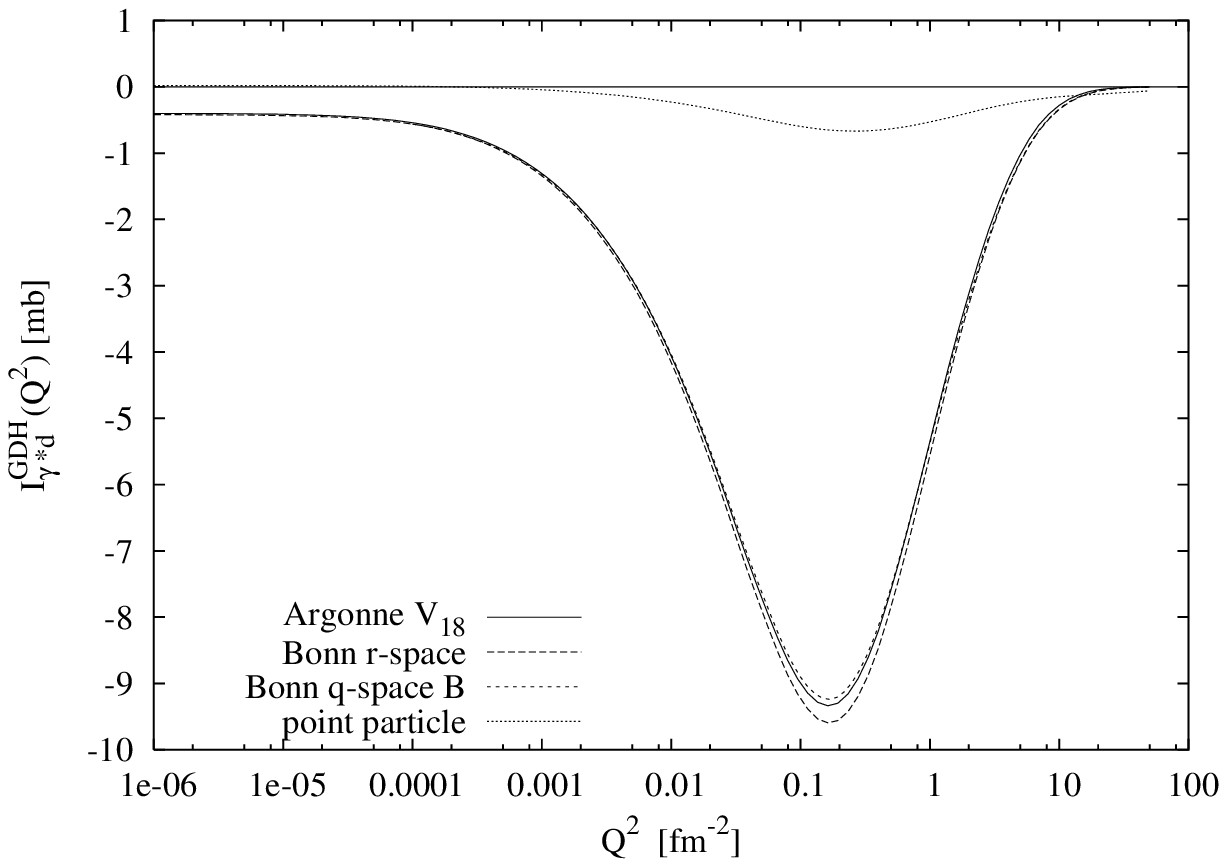}}
\vspace*{-.2cm}
\caption{Generalized Gerasimov-Drell-Hearn integral as function of
$Q^2$ for deuteron electrodisintegration $d(e,e')np$. Left panel:
separate current contributions from normal nonrelativistic theory (N)
and successively added meson exchange currents (MEC), isobar
configurations (IC), and relativistic contributions (RC). Right panel:
results of the complete calculation (T) 
for different potential models and for vanishing anomalous
nucleon magnetic moments (labeled ``point particle'').}
\vspace*{-.7cm}
\label{int_gdh_v18_log}
\end{figure}

Explicit evaluations of the generalized GDH integral for the 
electrodisintegration channel are 
exhibited in Fig.~\ref{int_gdh_v18_log}. The prominent feature is the
pronounced minimum around $Q^2\approx 0.2$~fm$^{-2}$ reflecting the 
absolutely largest spin asymmetry in Fig.~\ref{fig_gen_gdh1} for this 
value of $Q^2$. The left panel shows the influence of various 
interaction effects from MEC, IC and RC. Near the minimum, the largest 
interaction effect arises from MEC, increasing the depth by about 10~\%,
and to a smaller extent from IC while their influences in other 
regions of $Q^2$ is quite small. Relativistic contributions 
are substantial near the photon point as has been noted already for
photodisintegration\cite{ArK97}. But at higher $Q^2$ they are quite
tiny. The right panel of Fig.~\ref{int_gdh_v18_log} shows a comparison 
for three realistic potential models, the Bonn
r-space, the Bonn p-space (OBEPQ-B)\cite{MaH87} and the 
Argonne $V_{18}$\cite{WiS95}
models. Obviously, the potential model variation is quite small
compared to the interaction effects. The fact, that near threshold
the spin asymmetry is essentially determined by the nucleon isovector 
anomalous magnetic moments, is demonstrated by evaluating 
$I_{\gamma^* d}^{GDH}(Q^2)$ for vanishing
anomalous moments. The result is also shown in the right panel of
Fig.~\ref{int_gdh_v18_log} and is indeed quite tiny. Thus the disintegration 
contribution to the generalized GDH-integral is essentially 
driven by the nucleon's anomalous magnetic moments. 

\section{Conclusions and outlook}

\noindent
(i) Real photons:
\vspace*{.1cm}

\noindent
-- The spin asymmetry of the deuteron is a very 
interesting observable of its own value
because of a strong anticorrelation between  
low energy photodisintegration and 
at high energy meson production channels.\\
-- The spin asymmetry is very sensitive to relativistic effects at 
quite low energies which have never been tested in detail for this 
observable. \\
-- A direct access to the neutron spin asymmetry from the spin
asymmetry of the deuteron is not possible.\\ 
-- However, the deuteron spin asymmetry will provide a more detailed
test for $\pi$-production on the neutron and thus in an indirect 
manner on the spin asymmetry of the neutron.
\vspace*{.2cm}

\noindent
(ii) Virtual photons:
\vspace*{.1cm}

\noindent
-- The deuteron spin asymmetry of the electrodisintegration channel 
$d(e,e')np$ for $Q^2=$const.\ exhibits as function of 
the final state energy $E_{np}$ a pronounced minimum around 
$E_{np}\approx 70$~KeV, the location of the ``antibound''
$^1$S$_0$ state, which is deepest for $Q^2\approx 0.2$~fm$^{-2}$.\\ 
-- This minimum is dominated by a single M1 transition to this 
$^1$S$_0$ state and almost completely driven by the nucleon 
isovector anomalous magnetic moment.\\ 
-- An experimental check of this feature would provide a significant test
for our present theoretical understanding of the properties of few-body
nuclei.\\
-- The spin asymmetry falls off rapidly with increasing $E_{np}$ so that the
generalized GDH integral converges fast for this channel. \\
-- An independent check in the framework of effective field theory would be
very interesting.\\
-- As future task remains the evaluation of the other channels, like 
coherent und incoherent single and double pion and eta electroproduction.

\end{document}